\documentclass[prc,onecolumn,aps,floats,superscriptaddress,floatfix,
showpacs,showkeys,amsmath,amssymb]{revtex4}

\usepackage{dcolumn}
\usepackage{bm}
\usepackage{graphicx}
\usepackage{times}

\begin{document}
\title{Acoustic horizons in steady spherically symmetric 
nuclear fluid flows}
\author{Niladri Sarkar}\email{niladri.sarkar@saha.ac.in}
\author{Abhik Basu}\email{abhik.basu@saha.ac.in}
\affiliation{Condensed Matter Physics Division, Saha
Institute of Nuclear Physics, Calcutta 700064, India}
\author{Jayanta K. Bhattacharjee}\email{jkb@hri.res.in}
\affiliation{Harish Chandra Research Institute, Chhatnag Road,
Jhunsi, Allahabad 211019, India}
\author{Arnab K. Ray}\email{arnab.kumar@juet.ac.in}
\affiliation{Department of Physics, Jaypee University
of Engineering \& Technology, 
Raghogarh, Guna 473226, Madhya Pradesh, India}
\date{\today}

\begin{abstract}
We consider a hydrodynamic description of the spherically
symmetric outward flow of nuclear matter, using a nuclear 
model that introduces a weakly dispersive effect in the flow. 
About the resulting stationary conditions of the flow, we 
apply an Eulerian scheme to derive a fully nonlinear equation 
of a time-dependent radial perturbation. In its linearized limit,
with no dispersion, this equation implies the static acoustic
horizon of an analogue gravity model. This horizon also defines
the minimum radius of the steady flow.  
We model the perturbation as a high-frequency travelling
wave, in which the weak dispersion is taken iteratively. 
A {\it WKB} analysis shows that even arbitrarily small
values of dispersion make the horizon fully opaque to any
acoustic disturbance propagating against the bulk flow, with
the amplitude and the energy flux of the radial perturbation
decaying exponentially just outside the horizon. 
Nonlinear effects shift the horizon from its
steady position. 
\end{abstract}


\maketitle

\section{Introduction}
\label{sec1}

A wide variety of hydrodynamic flows, ranging from astrophysical
flows to flows in kitchen sinks, shows the existence of an
``acoustic 
horizon"~\cite{wgu81,vis98,su02,tkd04,vol05,dbd06,rr07,rb07,blv11,rob12}.
Such a horizon is attained
when the speed of the bulk flow matches the speed with which a
relevant wave carries information through the medium (for example,
sound waves or gravity waves). Passage of information is
uni-directional across the horizon, as it happens
analogously in the case of black holes or
white holes. This point of view, also known as analogue gravity,
leads to many interesting consequences in fluid flows. For
instance, in a low-dimensional flow, with viscosity lending an
additional effect, there is an abrupt increase
in the depth of the fluid at the horizon --- a phenomenon that is
popularly known as the hydraulic jump~\cite{wat64,bdp93,sbr05,rb07}.

In the distinctly different context of nuclear physics, in an 
energetic heavy-ion collision, a hot and expanding nuclear gas 
is formed~\cite{bgz78}. Hydrodynamic features are known to be 
collectively important in such cases, and appropriate hydrodynamic 
descriptions were furnished both for the expansion stage of the 
fluid~\cite{bgz78}, and its compression~\cite{cjtw73}. In the 
former case, an analytical treatment
was presented for the nonlinear hydrodynamic equations,
describing a free isentropic expansion~\cite{bgz78}.
The hydrodynamic equations invoked in this approach were the
familiar ones used in studies of radially outward flows in
spherical symmetry, and later studies were also to pursue the same
line~\cite{ccl98}. One of the early investigations at low energies
looked into the aftermath of nuclear collisions near the speed
of sound~\cite{frsw82}, in which the possibility of soliton
formation was considered, and so in due course dispersive terms
were
also required to be introduced in the hydrodynamic equations.
The subject of solitons in nuclear reactions was more formally
addressed in later works~\cite{rw83,hrw85}. Likewise, the
question of shock fronts was also accorded its rightful
importance~\cite{gr80}.

In nuclear fluids, hydrodynamic models with dispersion and
viscous effects have been a regular topic of study over the
last decade. A significant
amount of work including dispersive effects in particular,
has been carried out by now, all of which
required careful studies of suitable equations of state and
setting up of hydrodynamic equations, leading to solitonic
solutions in some cases~\cite{fona06,fona07,foffna10,fonaff11}.
And alongside dispersion, the importance of viscous hydrodynamics
in the description of nuclear matter at extreme energy densities
has not been overlooked
either~\cite{roro07,dt08,luro08,luro09,degv11}.

Given this overall background on hydrodynamics, in the present 
work we study the hydrodynamic aspects of nuclear matter from the
perspective of analogue gravity and acoustic horizons. For a 
radial outflow in spherical symmetry, we make use
of standard hydrodynamic model equations, 
accommodating a weakly dispersive effect that,  
nevertheless, has a strong influence on the flow. It is interesting
to note that the same influence may also be exerted by viscosity, 
which is a more natural attribute of a fluid (discussed at the 
end of Section~\ref{sec5}).  
We apply an Eulerian perturbation scheme on
a steady extended flow and, including nonlinearity to any 
arbitrary order, we obtain the proper form of the metric of an
acoustic horizon. We find that the acoustic horizon defines the 
minimum radius of the stationary flow. We also proceed to argue 
that nonlinearity has an adverse impact on the
analogy of a static acoustic horizon. In the linear limit, we
fashion the perturbation as a high-frequency travelling wave, and
see how small effects of dispersion influence the steady conditions.
Working iteratively by having recourse to the {\it WKB} method, 
we show that dispersion reduces the amplitude and the energy flux
of the radially propagating wave completely to zero. 

To summarize the principal results of our work, we have 
demonstrated the existence of an acoustic horizon, specifically
that of a white hole, in our chosen
model of nuclear hydrodynamics (Section~\ref{sec3}), with no 
physical flow solution admitted within the horizon 
(Section~\ref{sec4}). Linearized perturbations do not destabilize 
either the stationary flow or the horizon, and at the horizon of 
the white hole, all acoustic signals are fully extinguished due 
to the dispersive effect in the linear 
regime (Section~\ref{sec5} and Appendix~\ref{app2}). 
Nonlinearity, however, disturbs the precise condition of a 
stationary horizon (Section~\ref{sec3} and Appendix~\ref{app1}).      

\section{The hydrodynamic equations}
\label{sec2}

The hydrodynamic description that we have adopted here is
relevant to high-energy impacts or collisions, whose
result is an outflow
of the nuclear fluid~\cite{bgz78}. The outward flow is
described by a velocity field, $v$, and a baryonic density
field, $n$. The latter is related to the mass density,
$\rho$, by $\rho = Mn$, where $M$ is the
nucleon mass. The two fields, $v$ and $n$, are coupled
through two equations, one given by the condition of momentum
balance, and the other by the continuity equation.
These conditions are further supplemented by an equation of
state connecting $n$ to the local pressure, $P$. For a perfect
nuclear fluid, $P$ is related to the enthalpy per nucleon, $h$, 
under isentropic conditions, by~\cite{fona06}
\begin{equation}
\label{enthalpy}
{\boldsymbol \nabla P}=n {\boldsymbol \nabla} h,
\end{equation}
so that the condition for momentum balance can be set as
\begin{equation}
\label{euler}
\frac{\partial {\mathbf v}}{\partial t} +
\left({\mathbf v}\cdot \boldsymbol \nabla\right)
{\mathbf v} =-\frac{1}{M}{\boldsymbol \nabla} h.
\end{equation}
We now use a form of $h$, given as~\cite{fona06}
\begin{equation}
\label{enthal}
h=E\left(n_{\mathrm e}\right) +
\frac{Mc_{\mathrm s}^2}{2n_{\mathrm e}^2}
\left(3n^2 + n_{\mathrm e}^2-4nn_{\mathrm e}\right),
\end{equation}
in which, 
$c_{\mathrm s}$ is the speed of sound in the nuclear fluid
and $n_{\mathrm e}$ is the equilibrium density, about which
the energy per nucleon, $E(n)$, is expanded in a Taylor
series~\cite{fona06}. At this stage the hydrodynamics is 
both inviscid and non-dispersive. We can bring in a viscous 
term from the usual expressions of the stress tensor. To 
introduce a dispersive term, a specific model is required. 
Such a model~\cite{fona06} provides a dispersive term through 
the zeroth order of the Taylor expansion, augmenting  
$E(n_{\mathrm e})$ in Eq.~(\ref{enthal}) by a non-local 
term, something that can be done only in terms
of $n$, since $n_{\mathrm e}$ is spatially constant. 
This effectively amounts to crafting a small non-local 
effect about the usual baryon-vector meson local coupling,  
and this contribution is considered at the mean-field level. 
So with this understanding, we can write  
\begin{equation}
\label{eee}
E\left(n_{\mathrm e}\right)=
\left(\frac{g_V^2}{2m_V^2}\right)n_{\mathrm e} 
+ \frac{\chi \left(n_{\mathrm e}\right)}{n_{\mathrm e}}
+ \left(\frac{g_V^2}{m_V^4}\right)
\nabla^2 n. 
\end{equation}
Here $\chi (n_{\mathrm e})$ has a known form~\cite{fona06} and $g_V$ 
is the coupling constant of the baryon-vector meson interaction,
with $m_V$ being the mass of the vector meson field. 

Now with the help 
of Eqs.~(\ref{enthal})~and~(\ref{eee}), we can substitute $h$ 
in Eq.~(\ref{euler}), to obtain
\begin{widetext}
\begin{equation}
\label{mombal}
\frac{\partial {\mathbf v}}{\partial t}
+({\mathbf v}\cdot \boldsymbol \nabla){\mathbf v}
= -\frac{c_{\mathrm s}^2}{2n_{\mathrm e}^2}
\left(6n \boldsymbol \nabla n
-4n_{\mathrm e} \boldsymbol \nabla n  \right)
- \frac{g_V^2}{Mm_V^4} \boldsymbol \nabla \left(\nabla^2 n\right),
\end{equation}
\end{widetext}
which has the standard inertial and advective terms in the left
hand side. The first term in the 
right hand side involves the gradient of the baryonic density,
and is analogous to the pressure term in standard hydrodynamics.
The last term in Eq.~(\ref{mombal}) is the interaction term that
describes the non-local coupling between the baryons and the 
vector meson field. This term acts in the manner of a dispersion, 
and is of principal interest in our study. It was introduced in
nuclear hydrodynamics to investigate the formation and propagation
of solitons in nuclear matter~\cite{fona06}. We
are interested not so much in
propagating solitonic solutions, as we are in
knowing how this dispersion term affects, in a perturbative sense,
the stationary solution that is yielded by the rest of the terms
in Eq.~(\ref{mombal}). So we consider the dispersion term only
in the regime of extremely weak baryon-vector meson interactions.

While Eq.~(\ref{mombal}) gives one condition for the dynamics of
$v$ and $n$, another condition is also required. This is provided
by the continuity equation, going as
\begin{equation}
\label{cont}
\frac{\partial n}{\partial t}+ \boldsymbol \nabla \cdot
\left(n{\bf v}\right)=0,
\end{equation}
from whose stationary limit, we obtain the condition of
baryon conservation.

Considering now a spherically symmetric outward flow,
Eq.~(\ref{mombal}) is recast as
\begin{widetext}
\begin{equation}
\label{momrad}
\frac{\partial v}{\partial t}+ \frac{\partial}{\partial r}
\left[\frac{v^2}{2} + \frac{c_{\mathrm s}^2}{2n_{\mathrm e}^2}
\left(3n^2 - 4n_{\mathrm e}n \right)\right]
= - \zeta \frac{\partial}{\partial r}\left[\frac{1}{r^2}
\frac{\partial}{\partial r}\left(r^2\frac{\partial n}{\partial r}
\right)\right],
\end{equation}
\end{widetext}
in which, we have set $\zeta = g_V^2/Mm_V^4$ for notational
convenience, with $\zeta \longrightarrow 0$, when the
baryon-vector meson interaction is treated as just a
very small perturbative effect.
Likewise, in spherically symmetric geometry, Eq.~(\ref{cont}) is
rendered as
\begin{equation}
\label{contrad}
\frac{\partial n}{\partial t}+ \frac{1}{r^2}
\frac{\partial}{\partial r}\left(nvr^2 \right) = 0.
\end{equation}
Taken together, Eqs.~(\ref{momrad})~and~(\ref{contrad}) form a
closed set to describe the coupled dynamics of the fields,
$v(r,t)$ and $n(r,t)$. Our subsequent
analysis will be based on these two equations only.

\section{The perturbation and the acoustic horizon}
\label{sec3}

Under steady conditions and with $\zeta \longrightarrow 0$,
Eqs.~(\ref{momrad})~and~(\ref{contrad}) give the stationary 
fields, $v_0(r)$ and $n_0(r)$. About these stationary values, 
the perturbation schemes which we prescribe for both $v$ and $n$, 
respectively, are 
$v(r,t) = v_0(r) + v^\prime (r,t)$ and
$n(r,t) = n_0(r) + n^\prime (r,t)$, with the primed quantities 
representing time-dependent perturbations. Next, following an 
Eulerian perturbation treatment, adopted from the field of 
astrophysical accretion~\cite{pso80}, we define a new variable, 
$f=nvr^2$, whose steady value, $f_0$, is a constant,
governed by the condition,  
\begin{equation} 
\label{effnot} 
n_0 v_0 r^2 = f_0. 
\end{equation}
This fact can be verified easily from the stationary
form of Eq.~(\ref{contrad}), from which we can, therefore, extract
\begin{equation}
\label{npert}
\frac{\partial n^\prime}{\partial t}=
-\frac{1}{r^2}\frac{\partial f^\prime}{\partial r}.
\end{equation}
From the definition of $f$ itself, we further derive,
\begin{equation}
\label{fpert}
\frac{f^\prime}{f_0} = \frac{n^\prime}{n_0} +
\frac{v^\prime}{v_0}
+ \frac{n^\prime}{n_0}\frac{v^\prime}{v_0},
\end{equation}
in which we have maintained all admissible orders of nonlinearity.

Now, making use of Eq.~(\ref{npert}) in Eq.~(\ref{fpert}), we get
\begin{equation}
\label{vpert}
\frac{\partial v^\prime}{\partial t}=
\frac{v}{f}\frac{\partial f^\prime}{\partial t}
+ \frac{v^2}{f}\frac{\partial f^\prime}{\partial r}.
\end{equation}
It is worth stressing here that Eq.~(\ref{vpert}), which is fully
nonlinear, and Eq.~(\ref{npert}), together give a closed set of
conditions by which we can represent $n^\prime$
and $v^\prime$ exclusively in terms of $f^\prime$. We now need
an independent condition, on which we can
apply Eqs.~(\ref{npert})~and~(\ref{vpert}), and just such a
condition is afforded by Eq.~(\ref{momrad}). We take the
second-order time derivative of Eq.~(\ref{momrad}), and then on
it we apply the results implied by Eqs.~(\ref{npert})~and~(\ref{vpert}),
as well as the second-order time derivative of Eq.~(\ref{vpert}).
Consequently we obtain
\begin{widetext}
\begin{equation}
\label{perteq}
\frac{\partial}{\partial t}
\left(h^{tt}\frac{\partial f^\prime}{\partial t}\right) +
\frac{\partial}{\partial t}
\left(h^{tr}\frac{\partial f^\prime}{\partial r}\right) +
\frac{\partial}{\partial r}
\left(h^{rt}\frac{\partial f^\prime}{\partial t}\right) +
\frac{\partial}{\partial r}
\left(h^{rr}\frac{\partial f^\prime}{\partial r}\right) =
\zeta \frac{\partial}{\partial r}\left\{\frac{1}{r^2}
\frac{\partial}{\partial r}
\left[r^2\frac{\partial}{\partial r}
\left(\frac{1}{r^2}\frac{\partial f^\prime}{\partial r}\right)
\right]\right\},
\end{equation}
\end{widetext}
in which
\begin{equation}
\label{aitch}
h^{tt}=\frac{v}{f},\,\,\,h^{tr}=h^{rt}=\frac{v^2}{f},\,\,\,
h^{rr}=\frac{v}{f}\left(v^2 - a^2\right),
\end{equation}
with
\begin{equation}
\label{acous}
a^2= 3 c_{\mathrm s}^2 \frac{n}{n_{\mathrm e}}
\left(\frac{n}{n_{\mathrm e}}-\frac{2}{3}\right). 
\end{equation}
This last expression 
further suggests that there shall be no
acoustic propagation in the fluid if $n \leq 2n_{\mathrm e}/3$.
This limit on the steady value of the particle density defines
a low point, below which the flow loses the character of a fluid
continuum, with an acoustic propagation no longer possible.
The factor of $2/3$ is simply due
to the choice of the equation of state in Eq.~(\ref{momrad}).

It is particularly interesting to study Eq.~(\ref{perteq}) in
the limit of $\zeta=0$, i.e. when there is no non-local
baryon-vector meson
interaction. In this special case, not only do we get proper
background stationary solutions out of
Eqs.~(\ref{momrad})~and~(\ref{contrad}), but also,
going by the symmetry of
Eq.~(\ref{perteq}), we can recast it in a compact form as
\begin{equation}
\label{compact}
\partial_\mu \left(h^{\mu \nu}\partial_\nu
f^\prime \right) =0,
\end{equation}
with the Greek indices running from $0$ to $1$, under the
equivalence that $0$ stands for $t$ and $1$ stands for $r$. We 
see that Eq.~(\ref{compact}), or equivalently, Eq.~(\ref{perteq}),
is a nonlinear equation containing arbitrary orders of nonlinearity
in the perturbative expansion. 
If, however, we work with a linearized equation,
then $h^{\mu \nu}$, containing only the zeroth-order terms,
can be read from the matrix,
\begin{equation}
\label{matrix}
h^{\mu \nu }=\frac{v_0}{f_0}
\begin{pmatrix}
1 \hfill & v_0 \\
v_0  & v_0^2 - a_0^2 \hfill
\end{pmatrix},
\end{equation}
in which, $a_0 \equiv a_0(r)$, is the steady value of $a$.
A significant implication of the foregoing matrix
is that under steady conditions, an acoustic disturbance in the
fluid propagates with the speed, $a_0$, and its value is
determined when $n=n_0$ in Eq.~(\ref{acous}). 

Now, in Lorentzian geometry the d'Alembertian of a scalar field
in curved space is obtained from the metric, $g_{\mu \nu}$, as
\begin{equation}
\label{alem}
\Delta \varphi \equiv \frac{1}{\sqrt{-g}}
\partial_\mu \left({\sqrt{-g}}\, g^{\mu \nu}
\partial_\nu \varphi \right),
\end{equation}
where $g^{\mu \nu}$ is the inverse of the matrix,
$g_{\mu \nu}$~\citep{vis98,blv11}. We look for an equivalence
between $h^{\mu \nu }$ and $\sqrt{-g}\, g^{\mu \nu}$ by
comparing Eqs.~(\ref{compact})~and~(\ref{alem})
with each other, and we see that
Eq.~(\ref{compact}) gives an expression of $f^{\prime}$
that is of the type given by Eq.~(\ref{alem}). In the
linear order, the metrical part of Eq.~(\ref{compact}),
as Eq.~(\ref{matrix}) shows it, may then be extracted,
and its inverse will indicate the existence of an acoustic
horizon, when $v_0^2 = a_0^2$. In the case of a radially outflowing
nuclear fluid, this horizon is due to an acoustic white hole.
The radius of the horizon is the critical radius, $r_{\mathrm c}$,
that cannot be breached by any acoustic disturbance (carrying
any kind of information) propagating against the bulk outflow,
after having originated in the subcritical region,
where $v_0^2 < a_0^2$ and $r > r_{\mathrm c}$.
Effectively, then, we can say that the flow of information
across the acoustic horizon is uni-directional.

However, this discussion is valid only as far as the linear
ordering goes.
When nonlinearity is to be accounted for, then instead of
Eq.~(\ref{matrix}), it is Eq.~(\ref{aitch})
that defines the elements, $h^{\mu \nu}$, depending on
the order of nonlinearity that one wishes to retain (in principle
one could go up to any arbitrary order). The first serious
consequence of including nonlinearity is that the static 
description of $h^{\mu \nu}$, as stated in Eq.~(\ref{matrix}), 
will not suffice any longer. This view is in conformity with 
a numerical study~\cite{macmal08}, in which, for the case of 
spherically symmetric astrophysical accretion, it was shown 
that if the perturbations were to become strong, then acoustic 
horizons would suffer a radial shift about their previous 
static position, and the analogy between an acoustic horizon 
and the event horizon of a black hole (or a white hole) would 
appear limited. We have 
presented an analytical perspective of this argument in
Appendix~\ref{app1}.     

\section{The stationary background solutions}
\label{sec4}

Carrying forward the understanding that with $\zeta =0$, an
acoustic horizon may be expected when $v_0^2 =a_0^2$, we now
look for the stationary profile of the dispersion-free flow. 
To this end, apart
from the steady continuity condition, as given by Eq.~(\ref{effnot}), 
we also need the Bernoulli equation, obtained from the stationary 
limit of Eq.~(\ref{momrad}). This gives 
\begin{equation}
\label{bernou}
\frac{v_0^2}{2} + \frac{c_{\mathrm s}^2}{2n_{\mathrm e}^2}
\left(3n_0^2 - 4n_{\mathrm e}n_0\right) = E, 
\end{equation} 
with $E$ being the Bernoulli constant. It is expedient to work 
with dimensionless variables, under the scaling prescription, 
$x=n_0/n_{\mathrm e}$, $y=v_0/c_{\mathrm s}$, $u=a_0/c_{\mathrm s}$ 
and $R=r/\sqrt{f_0(c_{\mathrm s}n_{\mathrm e})^{-1}}$. In that 
case, Eqs.~(\ref{effnot})~and~(\ref{bernou}) can, respectively,
be recast as 
\begin{equation}
\label{scalef0}
xyR^2 =1
\end{equation}
and 
\begin{equation}
\label{scalebern}
y^2 + 3x^2 - 4x = B, 
\end{equation}
where, $B = 2E/c_{\mathrm s}^2$, is the scaled Bernoulli constant. 
The two foregoing equations provide a one-parameter family of
solutions, upon 
either retaining $y$ and eliminating $x$, or vice versa. 
We choose the former approach, because the condition of the acoustic 
horizon, which we already know to be $v_0^2 = a_0^2$, is better 
understood in terms of the velocity of the flow. So making use of 
the stationary function of $a_0^2$, as given by Eq.~(\ref{acous}),
which we scale as $u^2 = x(3x-2)$, we obtain 
\begin{equation}
\label{scalehor}
y_{\mathrm c}^2 = u_{\mathrm c}^2 = 
x_{\mathrm c}\left(3x_{\mathrm c}-2\right), 
\end{equation}
where we have used the subscript ``$\mathrm c$" to label all 
values at the horizon. The radius of the horizon, in the scaled
set of variables, is given by the condition 
$R_{\mathrm c}^2=(x_{\mathrm c}y_{\mathrm c})^{-1}$. With the 
help of Eq.~(\ref{scalehor}), we can, therefore, write,
\begin{equation}
\label{critrad}
R_{\mathrm c}^4=
\frac{1}{x_{\mathrm c}^3 \left(3x_{\mathrm c}-2\right)}. 
\end{equation} 
Our next task would be to show that $R_{\mathrm c}$ has a fixed
value. This can only happen when $x_{\mathrm c}$ has a fixed 
value, a condition that can be obtained by combining 
Eqs.~(\ref{scalebern})~and~(\ref{scalehor}), to yield  
\begin{equation}
\label{xquad} 
x_{\mathrm c}^2-x_{\mathrm c}-\left(B/6\right) = 0. 
\end{equation} 
With $x_{\mathrm c}$ having been fixed thus,  
$R_{\mathrm c}$ also becomes a 
fixed quantity. Keeping only the physically meaningful positive 
root of the discriminant of Eq.~(\ref{xquad}), we get 
$x_{\mathrm c}=[1+\sqrt{1+(2B/3)}]/2$. We note that 
at the horizon, $x_{\mathrm c}>1$, i.e. $n_0 >n_{\mathrm e}$.   

Now we would like to find the stationary velocity profile, 
$y \equiv y(R)$, which we obtain from 
Eqs.~(\ref{scalef0})~and~(\ref{scalebern}) as 
\begin{equation}
\label{velprof}
y^4 - By^2 - \frac{4y}{R^2} + \frac{3}{R^4} =0. 
\end{equation} 
What we have got is the equation of a quartic polynomial, a
mathematical condition that would have remained qualitatively 
unaltered if we had chosen the density profile, $x(R)$,
for our study. Before we solve for $y(R)$, it would be instructive
to look at the first derivative of $y$. This is given by 
\begin{equation}
\label{yderi1} 
\frac{{\mathrm d}y}{{\mathrm d}R} = 
\frac{2\left(3-2yR^2\right)}{R^3\left(2y^3R^2 -ByR^2 -2\right)}. 
\end{equation} 
The turning point in the velocity profile occurs when the numerator
in the right hand side of the first derivative vanishes. Making 
use of Eq.~(\ref{scalef0}), we find that the turning point 
corresponds to $x =2/3$, which, we recall to be the limiting 
value of $x$, below which the flow loses its continuum character. 
The singular point in the flow, when the denominator in the right
hand side of the derivative vanishes, offers a more interesting 
insight. In this case, going by 
Eqs.~(\ref{scalef0})~and~(\ref{scalebern}), we actually arrive 
at the condition, $x^2-x-(B/6)=0$, which is precisely the condition  
that prevails at the horizon, as given by Eq.~(\ref{xquad}). 
The simple conclusion to draw here is that the velocity profile 
becomes singular at the horizon. Further, taking the reciprocal 
of Eq.~(\ref{yderi1}) at the singular point, we get 
${\mathrm d}R/{\mathrm d}y=0$, which means that the horizon
is also the minimum radial position that flow solutions may 
reach. No solution is admitted within the horizon.  

Now, to know what the stationary velocity profile looks like, we
first have to view Eq.~(\ref{velprof}) in the standard form of 
a quartic equation, $y^4 + 2A_3y^3 + A_2y^2 + 2A_1y +A_0 =0$, 
noting the equivalence, $A_3 =0$, $A_2 =-B$, $A_1 =-2/R^2$ and
$A_0=3/R^4$. Evidently, Eq.~(\ref{velprof}) will yield four roots. 
These roots can be found analytically by using Ferrari's 
method of solving quartic equations. In order to do so, a term 
going like $(\Upsilon_1 y +\Upsilon_2)^2$ is to be added to both
sides of Eq.~(\ref{velprof}), and then the resulting left hand
side is required to be a perfect square in the form
$(y^2 +\kappa)^2$, so that the full equation will be rendered 
as $(y^2 +\kappa)^2=(\Upsilon_1 y +\Upsilon_2)^2$. This will 
deliver three conditions going as $\kappa^2 =A_0 +\Upsilon_2^2$, 
$\Upsilon_1 \Upsilon_2 = -A_1$ and $2\kappa =A_2 +\Upsilon_1^2$. 
Eliminating $\Upsilon_1$ and $\Upsilon_2$ from these three 
conditions, will deliver an auxiliary 
cubic equation in $\kappa$ going as 
$2\kappa^3 -A_2\kappa^2 -2A_0\kappa +(A_2A_0 -A_1^2)=0$, 
which, under the transformation, $\kappa = \Psi +(A_2/6)$, can 
be reduced to the canonical form of the cubic equation,
$\Psi^3 +P\Psi +Q=0$, with $P= -(A_2^2/12) - A_0$ and  
$Q= -(A_2^3/108) + (A_2A_0/3)-(A_1^2/2)$. 
Analytical solutions of the roots of the cubic 
equation in $\Psi$ can be obtained by the application of
the Cardano-Tartaglia-del Ferro method of solving cubic equations.
This will lead to the solution
\begin{equation}
\label{cardan}
\Psi = \left(-\frac{Q}{2} + \sqrt{\mathcal D} \right)^{1/3}
+ \left(-\frac{Q}{2} - \sqrt{\mathcal D} \right)^{1/3},
\end{equation}
with the discriminant, $\mathcal D$, having been defined by
${\mathcal D}=(Q^2/4)+(P^3/27)$. 
The sign of $\mathcal D$ is crucial here. If ${\mathcal D} >0$,
then there will be only one real root of $\Psi$, given directly by
Eq.~(\ref{cardan}). On the other hand, if ${\mathcal D} <0$,
then there will be three real roots of $\Psi$, all of which,
under a new definition,
$\vartheta =\arccos [-Q/\sqrt{-4(P/3)^3}]$, 
can be expressed in a slightly modified form as
\begin{equation}
\label{gamtheta}
\Psi_j = 2 \sqrt{\frac{-P}{3}} \cos \left[\frac{\vartheta + 2 \pi
\left(j -1 \right)}{3} \right],
\end{equation}
with the label, $j$, taking the values, $j=1,2,3$, 
for the three distinct roots. Of these three roots, the one 
corresponding to $j=1$ continues smoothly over to the root 
given by Eq.~(\ref{cardan}). So this root, $\Psi_1$, is the 
one of our choice. 
Once $\Psi$ is known thus, it is a simple task thereafter to find
$\kappa$, $\Upsilon_1$ and $\Upsilon_2$, all of which depend on 
$R$ and $B$. With this having been 
accomplished, all the four roots of $y$ in Eq.~(\ref{velprof})
can be obtained by solving the two distinct quadratic equations,
$y^2 +\kappa = \pm (\Upsilon_1 y + \Upsilon_2)$. Since we are 
concerned with a physical outflow with positive values of the
velocity profile, we choose the upper sign, and in consequence,
the two roots we get are 
\begin{equation} 
\label{velsol} 
y = \frac{1}{2}\left[\Upsilon_1 \pm 
\sqrt{\Upsilon_1^2 -4\left(\kappa -\Upsilon_2\right)}\right].  
\end{equation}
As soon as $y$ is obtained for a value of $R$, we can get $x(R)$
from Eq.~(\ref{scalef0}). Thereafter, $u(R)$ also becomes known. 
 
\begin{figure}[floatfix]
\begin{center}
\includegraphics[scale=1.0, angle=0]{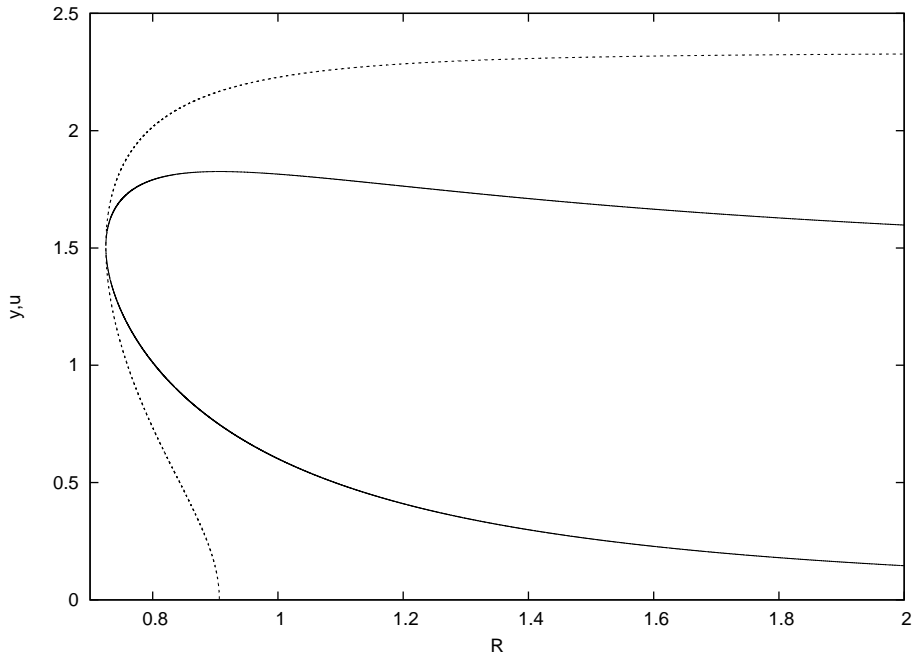}
\caption{\label{f1}\small{The plot of the velocity solution,
$y(R)$, as Eq.~(\ref{velsol}) gives it, is depicted by the
inner continuous curve (for $B=2$).
There are two branches of this curve, the upper one
corresponding to the positive sign of the discriminant in
Eq.~(\ref{velsol}) and the lower one corresponding
to the negative sign. Both branches meet at $y \simeq 1.5$,
where the flow becomes singular. The radial coordinate
of this point is the minimum radius of the flow, as
well as the radius of the acoustic horizon. There is a
maximum point in the upper branch, when $x=2/3$.
The outer envelope of a dotted curve shows the two branches
of $u(R)$, the speed of acoustic propagation. There are two
branches of this curve as well, both meeting $y(R)$ at the 
horizon. The upper flat branch of $u$ traces
the acoustic profile that goes with the lower branch of the
velocity profile, $y$. The lower branch of the acoustic profile,
going with the upper branch of $y$, vanishes abruptly when
$x=2/3$. This is the same position where the upper branch
of $y$ reaches its maximum. So, beyond this point, the upper
branch of $y$ loses the nature of a hydrodynamic continuum,
with no acoustic propagation being possible in it. The lower
branch of $y$ maintains its hydrodynamic features
globally, and decays asymptotically as $y \sim R^{-2}$.}}
\end{center}
\end{figure}
The overall behaviour of the stationary background flow is best 
comprehended from Fig.~\ref{f1}, in which, both $y(R)$ and $u(R)$ 
have been plotted. Either function has got two branches, 
corresponding to the two signs of the discriminant in 
Eq.~(\ref{velsol}). That there are two solutions of $y$ for every 
value of $R$, is actually a consequence of the invariance of 
Eqs.~(\ref{scalef0})~and~(\ref{scalebern}) under the 
transformation, $y \longrightarrow -y$. A similar symmetry 
exists in the mathematical problem of stationary 
accretion/wind~\cite{arc99}. Both the branches of $y$ and $u$ 
meet at $y=u \simeq 1.5$, where the flow becomes singular, and 
where the discriminant of Eq.~(\ref{velsol}) vanishes. 
The radial coordinate of this point is the minimum radius 
of the flow, and also, as we are already aware,   
the radius of the acoustic horizon. This effectively means
that the flow starts from the acoustic horizon itself, with 
no physical flow (solution of $y$) being able to penetrate 
the horizon. While
this may seem like a surprising result, precedence of it 
exists in the two-dimensional ideal shallow-water
flow~\cite{bdp93}. To understand this condition mathematically,
we need to realize that in Eq.~(\ref{scalebern}), there is no 
strong attractor term, dependent on the radial coordinate,
that will enable the flow to originate as deep as possible.   
In contrast, in stationary astrophysical accretion, 
gravity furnishes just such a term and makes the 
flow attain as small values of the radial coordinate as 
would be feasible under a given inner boundary 
condition~\cite{bon52,pso80,rb02}. One can also test that 
when gravity is ``switched off" in the momentum balance 
equation of the stationary accretion/wind problem, then  
the global flow features become qualitatively similar to 
what has been shown in Fig.~\ref{f1}, with velocity solutions 
of either the accretion branch or the wind branch surviving 
only outside
the acoustic horizon. Apropos of our present problem on nuclear
fluid flows, we conjecture that nuclear forces, which are 
strongly attractive on small length scales, could be a 
candidate mechanism for a stationary flow 
solution to physically breach the acoustic horizon.      

\section{Travelling waves and dispersion}
\label{sec5}

So far our discussion has been restricted to the case of $\zeta =0$. 
However, if the coupling term governing the non-local baryon-vector 
meson interaction is revived even to a very small extent, i.e.
when weak dispersion is accounted for
in Eq.~(\ref{perteq}), then the symmetric structure implicit
in Eq.~(\ref{compact}) will be disrupted.
Consequently, the precise condition of an acoustic horizon will
be lost. We note that this disruptive
effect is analogous to that caused by viscosity (dissipation) in
the flow for the case of the hydraulic jump~\cite{sbr05,rb07}, or
that due to the coupling of the flow with the geometry of spacetime
in general relativistic spherical accretion~\cite{ncbr07}. 
Nevertheless, the most important feature to emerge from the analogy
of a white hole horizon 
shall remain qualitatively unchanged, namely,
an acoustic disturbance propagating upstream from the sub-critical 
flow region, where $v_0^2 < a_0^2$, encounters an insurmountable 
obstacle when $v_0^2 = a_0^2$. 

We take up Eq.~(\ref{perteq}) in its linearized limit, which
means that all $h^{\mu \nu}$ are to be read 
from Eq.~(\ref{matrix}). However, now
we also account for dispersion in Eq.~(\ref{perteq}), with
$\zeta \neq 0$. We then treat the perturbation as a
high-frequency travelling wave, whose wavelength, $\lambda$, is
much less than the natural length scale in the fluid system,
$r_{\mathrm c}$, the radius of the acoustic horizon,
i.e. $\lambda \ll r_{\mathrm c}$. Usually linear perturbations 
do not destabilize the stationary background, 
especially when there are 
no source-like terms in the right hand side of Eq.~(\ref{perteq}). 
However, the merest 
presence of such terms gives rise to varying physical
behaviour of the linear perturbation. For instance, in the case 
of the shallow-water circular hydraulic jump, a viscosity-dependent
source term causes a large divergence in the amplitude of the 
perturbation in the vicinity of the horizon~\cite{rb07}. 
In contrast, the curvature of 
spacetime in general relativistic spherical accretion has 
a strongly stabilizing effect on the perturbation~\cite{ncbr07}.   
Along these lines, we now ask how the stationary 
background flow is affected by the dispersive source-like 
term in the right hand side of Eq.~(\ref{perteq}). 
The background flow of interest for us is represented by the 
lower branch of the stationary velocity function in Fig.~\ref{f1}. 
Hydrodynamic features are globally preserved along this branch,
starting from the horizon to arbitrarily large outer radii of the 
flow. Under these specifications, we use a solution of $f^\prime$, 
in which we separate spatial and temporal dependence by
\begin{equation}
\label{effsep}
f^\prime \left(r,t \right)
= \exp \left[i s\left(r\right)-i\omega t\right],
\end{equation}
with the understanding
that $\omega$ is much greater than any characteristic frequency
of the system. Applying the foregoing solution to Eq.~(\ref{perteq}),
and on multiplying the resulting expression throughout by
$(f_0/f^\prime )v_0^{-1}$, we arrive at
\begin{widetext}
\begin{eqnarray}
\label{sepsol}
& &-\omega^2 + 2 v_0 \omega \frac{{\mathrm d}s}{{\mathrm d}r}
+ \left(v_0^2 - a_0^2 \right)
\left[i\frac{{\mathrm d}^2 s}{{\mathrm d}r^2}
- \left(\frac{{\mathrm d}s}{{\mathrm d}r}\right)^2 \right]
- 2i\omega \frac{{\mathrm d}v_0}{{\mathrm d}r}
+ \frac{i}{v_0}
\frac{{\mathrm d}}{{\mathrm d}r}
\left[ v_0 \left(v_0^2 - a_0^2\right) \right]
\frac{{\mathrm d}s}{{\mathrm d}r} \nonumber \\
& &=\zeta n_0 \Bigg{\{}\left[i\frac{{\mathrm d}^4s}{{{\mathrm d}r^4}}
- 4 \frac{{\mathrm d}s}{{\mathrm d}r}
\frac{{\mathrm d}^3 s}{{\mathrm d}r^3}
-6i\left(\frac{{\mathrm d}s}{{\mathrm d}r}\right)^2
\frac{{\mathrm d}^2s}{{{\mathrm d}r^2}} -3
\left(\frac{{\mathrm d}^2s}{{{\mathrm d}r^2}}\right)^2
+ \left(\frac{{\mathrm d}s}{{{\mathrm d}r}}\right)^4\right] \nonumber \\
& & \qquad -\frac{4}{r}\left[i\frac{{\mathrm d}^3s}{{{\mathrm d}r^3}}
-3\frac{{\mathrm d}s}{{{\mathrm d}r}}
\frac{{\mathrm d}^2s}{{{\mathrm d}r^2}}
-i\left(\frac{{\mathrm d}s}{{{\mathrm d}r}}\right)^3 \right]
+\frac{8}{r^2}\left[i\frac{{\mathrm d}^2s}{{{\mathrm d}r^2}}
- \left(\frac{{\mathrm d}s}{{{\mathrm d}r}}\right)^2\right]
-i\frac{8}{r^3}\frac{{\mathrm d}s}{{{\mathrm d}r}}
\Bigg{\}}.
\end{eqnarray}
\end{widetext}
It is clear from Eqs.~(\ref{effsep})~and~(\ref{sepsol}) that $s$
should have both real and imaginary components. Therefore, we write
$s(r)=\alpha (r)+i\beta (r)$,
and viewing this solution along with Eq.~(\ref{effsep}), we can see
that while $\alpha$ contributes to the phase of the perturbation,
$\beta$ contributes to its amplitude.

Our approach to obtaining solutions of both $\alpha$ and $\beta$
is by the {\it WKB} analysis of Eq.~(\ref{sepsol}) for high-frequency
travelling waves. However, a look at Eq.~(\ref{sepsol}) also reveals
that the highest orders in it are quartic.
The saving grace is that all orders higher than
the second order, are dependent on $\zeta$, whose involvement has
been designed in our analysis to be very feeble anyway.
We exploit this fact by
first setting $\zeta =0$ in the right hand side of Eq.~(\ref{sepsol}),
and then we solve the
second-order differential equation implied
by the $\zeta$-independent left hand side of this equation.
To stress this special case, we also modify our solution as,
$s_0(r)= \alpha_0 (r)+i\beta_0 (r)$.
Using this in Eq.~(\ref{sepsol}), in which now $\zeta =0$, we
first separate the real and the imaginary parts, which are then
individually set equal to zero. The {\it WKB} prescription stipulates
that $\alpha_0 \gg \beta_0$. Going by this requirement, we, therefore,
collect only the real terms which do not contain $\beta_0$, and solve
a resulting quadratic equation
in ${\mathrm d}\alpha_0/{\mathrm d}r$ to obtain
\begin{equation}
\label{alphanot}
\alpha_0 = \int \frac{\omega}{{v_0 \mp a_0}}\,{\mathrm d}r.
\end{equation}
Likewise, from the imaginary part, in which we need to use the
solution of $\alpha_0$, we obtain
\begin{equation}
\label{betanot}
\beta_0 = \frac{1}{2} \ln \left( v_0 a_0 \right) + C,
\end{equation}
with $C$ being a constant of integration.

It should be pertinent now to perform a self-consistency check on
$\alpha_0$ and $\beta_0$, according to the condition of the {\it WKB}
analysis that $\alpha_0 \gg \beta_0$. First, we note that
with regard to the frequency, $\omega$, of the
high-frequency travelling waves, $\alpha_0$ (containing $\omega$)
is of a leading order over $\beta_0$ (containing $\omega^0$). Next,
on very large scales of length, i.e. $r \longrightarrow \infty$,
the background velocity goes asymptotically as
$v_0 \longrightarrow 0$, and the corresponding speed of acoustic 
propagation, $a_0$, approaches a
constant asymptotic value. In that case $\alpha_0 \sim \omega r$,
from Eq.~(\ref{alphanot}). Moreover, on similar scales of
length, going by Eqs.~(\ref{effnot})~and~(\ref{betanot}),
we see that $\beta_0 \sim \ln r$. Further, near the acoustic horizon,
where $v_0 \simeq a_0$, for the wave that goes against the bulk
flow with the speed, $v_0 - a_0$, we obtain 
large values of $\alpha_0$. All of these facts taken together, we
see that our solution scheme is very much in conformity with the
{\it WKB} prescription.

Thus far we have worked with $\zeta =0$ (absence of dispersion).
To know how dispersion affects the travelling wave, we now need to
find a solution of $s$ from Eq.~(\ref{sepsol}), with $\zeta \neq 0$.
To this end we adopt an iterative approach, with an imposition
of the condition
that $\zeta$ has a very small value. To put this in a properly
quantified perspective, we first find a scale of $\zeta$. A simple
dimensional analysis enables us to set this scale as
$\zeta_{\mathrm s}=c_{\mathrm s}^2 r_{\mathrm c}^2/n_{\mathrm e}$.
We then write $\zeta = \eta \zeta_{\mathrm s}$, where $\eta$ is a
dimensionless parameter that tunes the numerical value of $\zeta$.
In observance of the condition of our iterative method, we
require that $\eta \ll 1$.

Our next step is to take up Eq.~(\ref{sepsol}) in its full form
(with $\zeta \neq 0$ now),
and propose a solution for it as $s=s_0 + \delta s_1$, with
$\delta$ being another dimensionless parameter like $\eta$,
obeying the same requirement, i.e. $\delta \ll 1$.
Therefore, in the right hand side of Eq.~(\ref{sepsol}) all terms
which carry the product, $\eta \delta$, can be safely neglected
as being very small. This, in keeping with the principle of our
iterative treatment, will effectively mean that all the
surviving dispersion-related terms
in the right hand side of Eq.~(\ref{sepsol})
will go as $\eta s_0$. Further, by the {\it WKB} analysis,
we have also assured ourselves that $\alpha_0 \gg \beta_0$. So
we can afford to ignore all the $\beta_0$-dependent terms as well,
in the right hand side of Eq.~(\ref{sepsol}), when we compare
them with all the terms containing $\alpha_0$. And finally,
the most dominant $\alpha_0$-dependent real term to stand
out in the right hand side of Eq.~(\ref{sepsol}) is of the
fourth-degree, $({\mathrm d}\alpha_0/{\mathrm d}r)^4$.
Preserving only this term in the right hand side of Eq.~(\ref{sepsol})
and extracting only the $\beta$-independent real terms from the
left hand side, we are left with a simple quadratic equation,
\begin{widetext}
\begin{equation}
\label{alphaiter}
\left(v_0^2 -a_0^2 \right)
\left(\frac{{{\mathrm d}\alpha}}{{{\mathrm d}r}}\right)^2
- 2 v_0 \omega
\frac{{{\mathrm d}\alpha}}{{{\mathrm d}r}}
+ \left[\omega^2 + \eta \zeta_{\mathrm s} n_0
\left(\frac{{{\mathrm d}\alpha_0}}{{{\mathrm d}r}}\right)^4
\right] =0,
\end{equation}
\end{widetext}
to solve for $\alpha$. Under the provision of
$\eta \ll (\lambda/r_{\mathrm c})^2$, which is well
in accordance with our requirement that $\eta$ may be arbitrarily
small, we obtain a solution of $\alpha$, going as
\begin{equation}
\label{alphacor}
\alpha \simeq \alpha_0 \mp \frac{1}{2} \eta \zeta_{\mathrm s}
\omega^3 \int \frac{n_0}{{a_0 \left(v_0 \mp a_0\right)^4}}\,
{\mathrm d}r.
\end{equation}

Similarly, to solve for $\beta$, we extract all the imaginary
terms from the left hand side of Eq.~(\ref{sepsol}), and noting
that the most dominant contribution to the imaginary terms in
the right hand side comes from the cubic-order terms involving
$\alpha_0$, we are required to solve the equation,
\begin{widetext}
\begin{equation}
\label{betaiter}
2 \left[v_0 \omega -\left(v_0^2 -a_0^2\right)
\frac{{\mathrm d}\alpha}{{\mathrm d}r}\right]
\frac{{\mathrm d}\beta}{{\mathrm d}r} -2\omega
\frac{{\mathrm d}v_0}{{\mathrm d}r} + \frac{1}{v_0}
\frac{\mathrm d}{{\mathrm d}r}
\left[v_0 \left(v_0^2 -a_0^2\right)
\frac{{\mathrm d}\alpha}{{\mathrm d}r} \right]
= 4 \eta \zeta_{\mathrm s}\frac{n_0}{r}
\left(\frac{{\mathrm d}\alpha_0}{{\mathrm d}r}\right)^3
\left\{1-\frac{3}{2}
\frac{{\mathrm d} \left[\ln \left({\mathrm d}\alpha_0/
{\mathrm d}r\right) \right]}
{{\mathrm d}\left(\ln r\right)}
\right\},
\end{equation}
\end{widetext}
from which, on using Eq.~(\ref{alphacor}), once again
under the condition
that $\eta \ll (\lambda/r_{\mathrm c})^2$, we obtain
\begin{equation}
\label{betacor} \beta \simeq \beta_0 - \eta \zeta_{\mathrm s}
\omega^2 {\mathcal H}\left(r\right),
\end{equation}
with $\mathcal H$ to be expressed fully as
\begin{widetext}
\begin{equation}
\label{hdisper}
{\mathcal H}\left(r\right) = \frac{1}{4}\frac{n_0}{a_0^2}
\left[\frac{v_0^2 -a_0^2}{\left(v_0 \mp a_0\right)^4}\right]
\mp 2 \int
\frac{n_0}{ra_0 \left(v_0 \mp a_0\right)^3}
\left\{1+\frac{3}{2}
\frac{{\mathrm d} \left[\ln \left(v_0 \mp a_0\right)\right]}
{{\mathrm d}\left(\ln r\right)}
\right\}\, {\mathrm d}r.
\end{equation}
\end{widetext}
The significant aspects of both
Eqs.~(\ref{alphacor})~and~(\ref{betacor}) are that in the former, 
the correction to the zero-dispersion condition is of the order 
of $\omega^3$ (an odd order contributing to the phase), and in the
latter a similar correction is of the order of $\omega^2$ (an even
order contributing to the amplitude). Now, noting that $\alpha_0$ 
is of the order of $\omega$ and $\beta_0$ is of the order of
$\omega^0$, the corrections to the zero-dispersion terms in both
Eqs.~(\ref{alphacor})~and~(\ref{betacor}) appear, in the
high-frequency regime, to be dominant over their respective zeroth
orders. This, however, is not really the case, as we shall 
argue. We have obtained the results given by
Eqs.~(\ref{alphacor})~and~(\ref{betacor}) under the restriction 
that $\eta \ll (\lambda/r_{\mathrm c})^2$. Once we view the
wavelength, $\lambda$, as $\lambda (r) = 2 \pi (v_0
\mp a_0)/\omega$, we immediately see that the combination of
$\eta \omega^2$ in Eqs.~(\ref{alphacor})~and~(\ref{betacor}),
reduces both the correction terms on $\alpha_0$ and $\beta_0$ 
to be sub-leading to their respective zero-order terms. While 
this appears to be true over most of the spatial range of the 
flow, an exception is to be made 
in the close neighbourhood of the acoustic horizon, where 
$v_0=a_0$. In this region, looking at Eq.~(\ref{betacor}) in
particular, we see that the correction on $\beta_0$ diverges, 
while $\beta_0$ itself remains finite. Hereafter, 
we are interested primarily in the correction on $\beta_0$.

Going back to Eq.~(\ref{effsep}), we are able to write,
$f^\prime (r,t) = e^{-\beta}
\exp (i\alpha -i\omega t)$, from which, by extracting
the amplitude part only, and also making use
of Eqs.~(\ref{betanot})~and~(\ref{betacor}), we get
\begin{equation}
\label{ampli}
\vert f^\prime \left(r,t\right) \vert
\simeq \frac{\tilde{C}}{\sqrt{v_0 a_0}}
\exp \left[\eta \zeta_{\mathrm s} \omega^2 {\mathcal H}\left(r\right)
\right],
\end{equation}
where $\tilde{C}$ is a constant. The influence of dispersion on the
amplitude can be seen from Eq.~(\ref{hdisper}), which, for a wave
moving upstream against the bulk outward 
flow, indicates a divergence at the
acoustic horizon, where $v_0 = a_0$. However, a careful examination
shows that if the wave moves towards the acoustic horizon from the
sub-critical region, where $v_0 < a_0$, then the divergence in
$\mathcal{H}$ carries a negative sign. Raised to an exponent, as
indicated by Eq.~(\ref{ampli}), this will mean that the amplitude 
of the wave will decay to zero at the acoustic horizon, 
which behaves like an impervious wall to any signal
approaching it from the sub-critical region. The energy flux of 
the perturbation also behaves in a manner similar to its amplitude, 
as we have shown in Appendix~\ref{app2}. 
Contrary to the wave travelling inwards, 
the solution of the perturbation travelling outwards 
with the bulk flow, must originate at or just outside
the horizon itself, since no steady background solutions are 
admitted within the horizon. So, taking all these facts together, 
we say that the passage 
of information at the horizon is uni-directional, and the horizon 
may be viewed analogously as a white hole. 
This point of view is, however, different from the analogue of 
the Hawking radiation in the problem of an acoustic metric, in 
which, without the kind of dispersive correction that we have 
used here, an acoustic signal is allowed to cross the 
acoustic horizon with a finite and non-zero (but spatially 
decreasing) amplitude~\cite{wgu81,tkd04,rob12}. At the horizon
of a general relativistic black hole, a similar feature is seen
on carrying out a {\it WKB}-type analysis of incoming and outgoing 
probability amplitudes~\cite{pm07}.  
In contrast, in our study, not only does the amplitude of the
perturbation drop sharply to zero at the horizon, making the 
horizon fully opaque to acoustic signals, but also physical 
flow solutions meet a dead end at an infinitely rigid horizon 
surface. Going by the former observation,  
our system is more akin to a classical
black hole without any analogue of the Hawking radiation.

In passing we also examine the possibilities presented by 
viscosity, although it has a very feeble presence in the 
type of nuclear fluid that we are studying. Nevertheless, if 
we had included viscous effects in our study, then the right 
hand side of Eq.~(\ref{mombal}) would have contained terms like 
$\eta^\star \nabla^2 {\mathbf v}$ and 
$[(\eta^\star/3)+\zeta^\star] \boldsymbol \nabla (\boldsymbol \nabla
\cdot {\mathbf v})$, where $\eta^\star$ and $\zeta^\star$ are the 
first and second coefficients of viscosity, respectively~\cite{ll87}. 
For the case of a compressible, irrotational and spherically symmetric
flow, the differential operators in the two viscosity terms
assume identical forms, bearing only the radial variation of the 
velocity field, $v(r,t)$~\cite{ray03}. Taking the time derivative
of these terms, and then making use of Eq.~(\ref{vpert}) in the 
context of our perturbation scheme, will involve viscosity in the  
field equation of $f^\prime (r,t)$, with the stationary coefficients 
of the viscosity terms containing second-order spatial derivatives 
of $v_0(r)$. Noting that near the horizon, 
${\mathrm d}v_0/{\mathrm d}r$ approaches very high values, the 
viscosity-dependent terms, making dominant contributions to the 
perturbation, shall emerge with ${\mathrm d}^2 v_0/{\mathrm d}r^2$
and $({\mathrm d}v_0/{\mathrm d}r)^2$. This will be similar to 
the dispersion-related term in the right hand side of 
Eq.~(\ref{betaiter}), and so we can say that viscosity will enter
the amplitude of the perturbation at the same order as dispersion.     

\section{Concluding Remarks}
\label{sec6}

In this work we have studied nuclear fluids from a hydrodynamic
perspective, with baryon-vector meson interactions bringing
dispersion as a novelty to the standard hydrodynamics. Not 
accounting for dispersion, the hydrodynamic flow yields an 
analogue metric, in the likeness of what is seen for a scalar 
field in curved spacetime. Extending this point of view, we have
found a critical horizon in the flow of a nuclear fluid, 
a feature that is reminiscent of a white hole. So, for a nuclear 
fluid flowing radially outwards, the horizon will be an opaque 
barrier to a wave propagating radially inwards through the fluid. 
This effect becomes particularly pronounced, when dispersion 
forces the amplitude of an acoustic signal
to suffer a much stronger decay than what the simple
hydrodynamics might admit, and this is in stark 
contrast to the Hawking radiation in an analogue black hole,
where the horizon is not entirely opaque to an outgoing signal
originating within the horizon. 

With dispersion incorporated, however, the symmetric form of an 
analogue metric is lost. This is similar to the way in which the 
coupling of the flow and the geometry of Schwarzschild spacetime
adversely affects the clear-cut horizon condition obtained otherwise
in the Newtonian construct of space and time~\cite{ncbr07}.
Qualitatively speaking, the same behaviour is also exhibited by
viscous dissipation~\cite{rb07}.
The influence of dispersive effects on the amplitude of
high-frequency travelling waves can also be compared to
the way viscous dissipation can act under similar
circumstances. In the low-dimensional problem of the
shallow-water hydraulic jump, viscosity is known to enhance
the amplitude of a high-frequency travelling wave, as it
moves against a radially outward bulk flow and approaches
the acoustic horizon of an analogue white hole from
the sub-critical flow region~\cite{rb07}. This is completely 
contrary to the way in which dispersion decays the
amplitude of a wave that arrives at the acoustic horizon
from the sub-critical region. One way
or the other, we realize now that dissipation~\cite{rb07},
the geometry of curved spacetime~\cite{ncbr07},
nonlinearity~\cite{macmal08} and dispersion, all appear to 
disturb the symmetric structure of an acoustic metric. 

An integral aspect of hydrodynamics is the equation of state, 
by which the pressure term in the momentum balance condition 
is closed. Depending on the nature of physical problems, the 
equation of state is prescribed variously. For instance, in 
astrophysical fluids, the standard formula is 
polytropic~\cite{bon52}, while in the shallow-water hydraulic 
jump, a linear equation of state is applied~\cite{bdp93}.    
The equation of state that we have used, specialized for our 
study of a nuclear fluid flow~\cite{fona06}, 
is given by a composite function, 
bearing a linear term and a second-order term of the baryonic
density. Hydrodynamic features hold true when the latter term
is effective, and break down under the dominance of the linear
term. This is an unusual physical aspect designed into this 
problem only, and is not to be seen when the equation of state 
is set by a single power-law term, as is usually the case.   
In a more general sense, however, as long as the equation 
of state, regardless of its particular form, provides a  
condition for an acoustic propagation, the mathematical 
procedure leading to an analogue metric and an acoustic
horizon remains universal.   

\begin{acknowledgments}
AB gratefully acknowledges partial financial support in the form 
of the Max-Planck Partner Group at the Saha Institute of Nuclear
Physics, Calcutta, India, funded jointly by the 
Max-Planck-Gesellschaft (Germany) and the Department of Science 
and Technology (India) through the Partner Group programme (2009). 
Comments from an anonymous referee have helped in improving this
work. 
\end{acknowledgments}

\appendix

\section{Nonlinearity and the acoustic horizon}
\label{app1} 

At the end of Section~\ref{sec3} we indicated that the 
acoustic horizon would get displaced under nonlinear effects. 
We can now verify this contention analytically by confining our
mathematical treatment to the second order of nonlinearity.
All of the nonlinearity in Eq.~(\ref{compact}) is contained 
in the metric elements, $h^{\mu \nu}$, involving the exact 
field variables, $v$, $a$ and $f$, as opposed to containing
only their stationary background counterparts~\cite{vis98,su02}.
This is indeed going into the realm of nonlinearity,
because $v$ and $a$ depend on $f$, while
$f$ is related to $f^\prime$ as $f=f_0+f^\prime$.

If we restrict ourselves to the second order of nonlinearity,  
we see that $h^{\mu \nu}$ in Eq.~(\ref{aitch}) will
bear primed quantities in their first power only.
Taken together with Eq.~(\ref{perteq}), this will, in effect, 
preserve all the terms which are nonlinear in the second
order. So we carry out the necessary expansion of
$v=v_0+v^\prime$, $n = n_0 +n^\prime$ and
$f=f_0+f^\prime$ in Eq.~(\ref{aitch}) up to
the first order only. The next process to perform
is to substitute both $n^\prime$ and $v^\prime$
with $f^\prime$. To make this substitution possible,
first we make use of Eq.~(\ref{fpert})
to represent $v^\prime$ in terms of $n^\prime$ and
$f^\prime$. While doing so, the
product term of $n^\prime$ and $v^\prime$ in
Eq.~(\ref{fpert}) is to be ignored, because
including it will bring in
the third order of nonlinearity. Once $v^\prime$ has
been substituted in this manner,
we have to write $n^\prime$ in terms of $f^\prime$.
This can be done by invoking the linear relationship
suggested by Eq.~(\ref{npert}),
with the reasoning that if $n^\prime$ and $f^\prime$
are both separable functions of space and time, with the
time part being exponential (all of which are standard
mathematical prescriptions in perturbative analysis), then
\begin{equation}
\label{nflin}
\frac{n^\prime}{n_0} = \sigma \left(r \right)
\frac{f^\prime}{f_0},
\end{equation}
with $\sigma$ being a function of $r$ only.
We shall self-consistently support this mathematical
argument in Appendix~\ref{app2}.
The exact functional form of $\sigma$ will
be determined by the way the spatial part of $f^\prime$ is
set up, but when $n^\prime$, $v^\prime$ and $f^\prime$
are all real fluctuations, $\sigma$ should likewise be real.

Following all of these tedious but straightforward algebraic
details, the non-stationary features of the elements, $h^{\mu \nu}$,
in Eq.~(\ref{compact}),
can finally be expressed entirely in terms of $f^\prime$ as
\begin{widetext}
\begin{equation}
\label{aitch2}
h^{tt} \simeq \frac{v_0}{f_0} \left(1 +\epsilon \xi^{tt}
\frac{f^\prime}{f_0}\right),\,\,\,
h^{tr} \simeq \frac{v_0^2}{f_0} \left(1 +\epsilon \xi^{tr}
\frac{f^\prime}{f_0}\right),\,\,\,
h^{rt} \simeq \frac{v_0^2}{f_0} \left(1 +\epsilon \xi^{rt}
\frac{f^\prime}{f_0}\right),\,\,\,
h^{rr} \simeq \frac{v_0}{f_0} \left(v_0^2-a_0^2\right)+\epsilon
\frac{v_0^3}{f_0} \xi^{rr} \frac{f^\prime}{f_0},
\end{equation}
\end{widetext}
in all of which, $\epsilon$ has been introduced as a nonlinear
``switch" parameter to keep track of all the nonlinear terms.
When $\epsilon =0$, only linearity remains. In this
limit we return to the familiar linear result implied
by Eq.~(\ref{matrix}). In the opposite extreme, when
$\epsilon =1$, in addition to the linear effects, the second
order of nonlinearity becomes activated
in Eq.~(\ref{compact}), and the stationary
position of an acoustic horizon gets disturbed due to the
nonlinear $\epsilon$-dependent terms. A look at $h^{rr}$
in Eq.~(\ref{aitch2}) makes this fact evident. We also note
that Eq.~(\ref{aitch2}) contains
the factors, $\xi^{\mu \nu}$, which are to be read as
\begin{widetext}
\begin{equation}
\label{zyees}
\xi^{tt}= -\sigma,\,\,\,
\xi^{tr}=\xi^{rt}=1-2\sigma,\,\,\,
\xi^{rr}= 2 - 3 \sigma \left[1+
\left(\frac{n_0}{n_{\mathrm e}}\right)^2
\left(\frac{c_{\mathrm s}}{v_0}\right)^2\right], 
\end{equation}
\end{widetext}
and all of which are just stationary quantities. 

\section{Energy flux of the travelling wave}
\label{app2} 

What we have seen as regards the amplitude of the travelling
wave in Section~\ref{sec5}, can also be seen in the energy flux 
associated with the travelling perturbation. We note first that 
the kinetic energy per unit volume of the flow is
\begin{equation}
\label{ekin}
{\mathcal{E}}_{\mathrm{kin}}=
\frac{1}{2} M \left(n_0 + n^\prime \right)
\left(v_0 + v^\prime \right)^2,
\end{equation}
and the internal energy per unit volume is
\begin{equation}
\label{eint}
{\mathcal{E}}_{\mathrm{int}}=
M \left[n_0 \varepsilon + n^\prime \frac{\partial}{\partial n_0}
\left(n_0 \varepsilon \right) + \frac{{n^\prime}^2}{2}
\frac{\partial^2}{\partial n_0^2}\left(n_0 \varepsilon \right)\right],
\end{equation}
where $\varepsilon$ is the internal energy per unit mass~\cite{ll87}.
In both of the foregoing expressions of energy, the zeroth-order
terms refer to the background flow, and the first-order terms
can be made to disappear upon performing a time averaging.
Thereafter, the time-averaged total energy in the perturbation,
per unit volume of fluid, 
is to be obtained by summing the second-order terms in
${\mathcal{E}}_{\mathrm{kin}}$ and ${\mathcal{E}}_{\mathrm{int}}$.
All of these terms will go either as ${n^\prime}^2$ or
${v^\prime}^2$, or a product of $n^\prime$ and $v^\prime$. Our
next task would be to represent both $n^\prime$ and $v^\prime$
in terms of $f^\prime$, and then use Eq.~(\ref{ampli}) to
substitute $f^\prime$. For this purpose, making use of
Eqs.~(\ref{npert}),~(\ref{effsep})~and~(\ref{alphanot}), we get
\begin{equation}
\label{eneff}
\frac{n^\prime}{n_0} \simeq \frac{v_0}{v_0 \mp a_0}
\frac{f^\prime}{f_0}.
\end{equation}
We can see now that this result
is precisely what Eq.~(\ref{nflin}) implies. Going
further, we make use of Eq.~(\ref{eneff}) in Eq.~(\ref{fpert}),
ignoring the product of $n^\prime$ and $v^\prime$ in the latter,
to obtain
\begin{equation}
\label{veff}
\frac{v^\prime}{v_0} \simeq \mp \frac{a_0}{v_0 \mp a_0}
\frac{f^\prime}{f_0}.
\end{equation}
Once we have two relations explicitly connecting $n^\prime$
and $v^\prime$ with $f^\prime$, as implied by
Eqs.~(\ref{eneff})~and~(\ref{veff}), we can derive
the time-averaged total energy per unit volume as
\begin{equation}
\label{etot}
{\mathcal{E}}_{\mathrm{tot}} =
\frac{M n_0 v_0^2 a_0^2}{2\left(v_0 \mp a_0\right)^2}
\left[1 \mp 2 \frac{v_0}{a_0}
+ \frac{n_0}{a_0^2}
\frac{\partial^2 \left(n_0 \varepsilon\right)}{\partial n_0^2}
\right] \frac{\langle {f^\prime}^2 \rangle}{f_0^2}.
\end{equation}
The energy flux, $\mathcal F$, associated with the
spherical wavefront, travelling with the speed,
$(v_0 \mp a_0)$, is
${\mathcal F} = 4 \pi r^2 {\mathcal E}_{\mathrm{tot}}
(v_0 \mp a_0)$. Accounting for a factor of $1/2$
due to the time-averaging of the phase part of ${f^\prime}^2$,
the flux can be written as
\begin{widetext}
\begin{equation}
\label{flux}
{\mathcal F} = \frac{2 \pi {\tilde C}^2 M}{f_0}
\left\{ \mp 1 - \frac{a_0}{2\left(v_0 \mp a_0\right)}
\left[1 - \frac{n_0}{a_0^2}
\frac{\partial^2 \left(n_0 \varepsilon\right)}{\partial n_0^2}
\right] \right\}
\exp \left[2 \eta \zeta_{\mathrm s}\omega^2
{\mathcal H}\left(r\right) \right].
\end{equation}
\end{widetext}
We realize immediately that near the acoustic horizon, the
dispersion-dependent exponential factor in Eq.~(\ref{flux})
will have very much the same effect on the energy
flux of the perturbation, as it has on its amplitude,
$\vert f^\prime \vert$, given by Eq.~(\ref{ampli}).

\bibliography{prc_sbbr2013r2}

\end{document}